\documentclass[pra,a4paper]{revtex4}
\usepackage{amsmath} 
\usepackage{amsfonts}
\usepackage{amssymb}
\usepackage{bm} 
\usepackage{graphicx}
\usepackage[applemac]{inputenc}
\usepackage[colorlinks]{hyperref}
\usepackage{epstopdf}

\begin{document}
\title{Comment on ``Ballistic SNS sandwich as a Josephson junction''}
\author{Erkki Thuneberg}
\affiliation{QTF Centre of Excellence, Department of Applied Physics, Aalto University, FI-00076 Aalto, Finland}
\date{\today} 
\begin{abstract}
The old problem of supercurrent in a long ballistic SNS Josephson junction is interpreted using a simple picture of the fermion energy levels in the normal and superconducting regions.  We argue that a recent paper on the topic by Sonin contains erroneous results. 
\end{abstract}

\maketitle

In a recent paper Sonin reports new results for one-dimensional ballistic SNS Josephson junction \cite{Sonin21}. 
This paper has rather limited experimental applicability since it neglects all scattering mechanism besides Andreev scattering. However, the paper discusses the important theoretical problem of  how to calculate the current in a degenerate Fermi system. Therefore we feel that it is important to point out that the main results of the paper are incorrect. Besides this, our central purpose is to explain the correct results dating from 1970'  in a new way, which hopefully leads to better understanding of the physics.

One argument against Sonin's results is that they are contrary to a wast literature on the topic. In particular, the ballistic SNS junction in equilibrium has been studied in references  \cite{Ishii70,Svidzinsky71,Bardeen72, Ishii72,Svidzinsky73,Bezuglyi75,Kupriyanov81,Furusaki91,Golubov04}. For completeness, we include one more calculation in the Appendix. In addition,  there is extensive literature on various generalizations to include normal scattering, dynamics and different  types superconductors and geometries. Note also that the papers by Svidzinskii et al.\ especially warn about the method of summing separately the discrete and continuum part of the spectrum -- the method used in Sonin's paper and in the original work by Kulik \cite{Kulik69}.

A reliable method to solve many problems in superconductivity and superfluidity is to use the quasiclassical Green's function formalism. This theory is thoroughly discussed by Serene and Rainer \cite{Serene83} in connection of superfluid $^3$He. This theory gives a straightforward procedure to solve various problems in superconductivity, some of which are discussed in Ref.\ \cite{Belzig99}. In this Comment we try to interpret the basic ideas and results in a simple way, in line with our earlier work \cite{Thuneberg06,KuorelahtiLaine18}. The mathematical interpretation is explained in the Appendix.

For simplicity of presentation, we make the same assumptions about the SNS junction as in Ref.\  \cite{Sonin21}. The order parameter $\Delta(x)$ vanishes in the normal region of length $L$, which is long compared to the superconductor coherence length. The order parameter has constant values $\Delta e^{\pm i\phi/2}$ in  the two superconducting regions, where $\phi$ is the phase difference over the junction. Ordinary  scattering in the normal and superconducting regions  as well at the NS interfaces is neglected. In addition, we assume zero temperature, $T=0$, and phase difference not exceeding the critical one, $|\phi|<\pi$. Not all assumptions are necessary for the main conclusions, and some generalizations are  discussion below. 

\begin{figure}[tb] 
   \centering
   \includegraphics[width=0.55\linewidth]{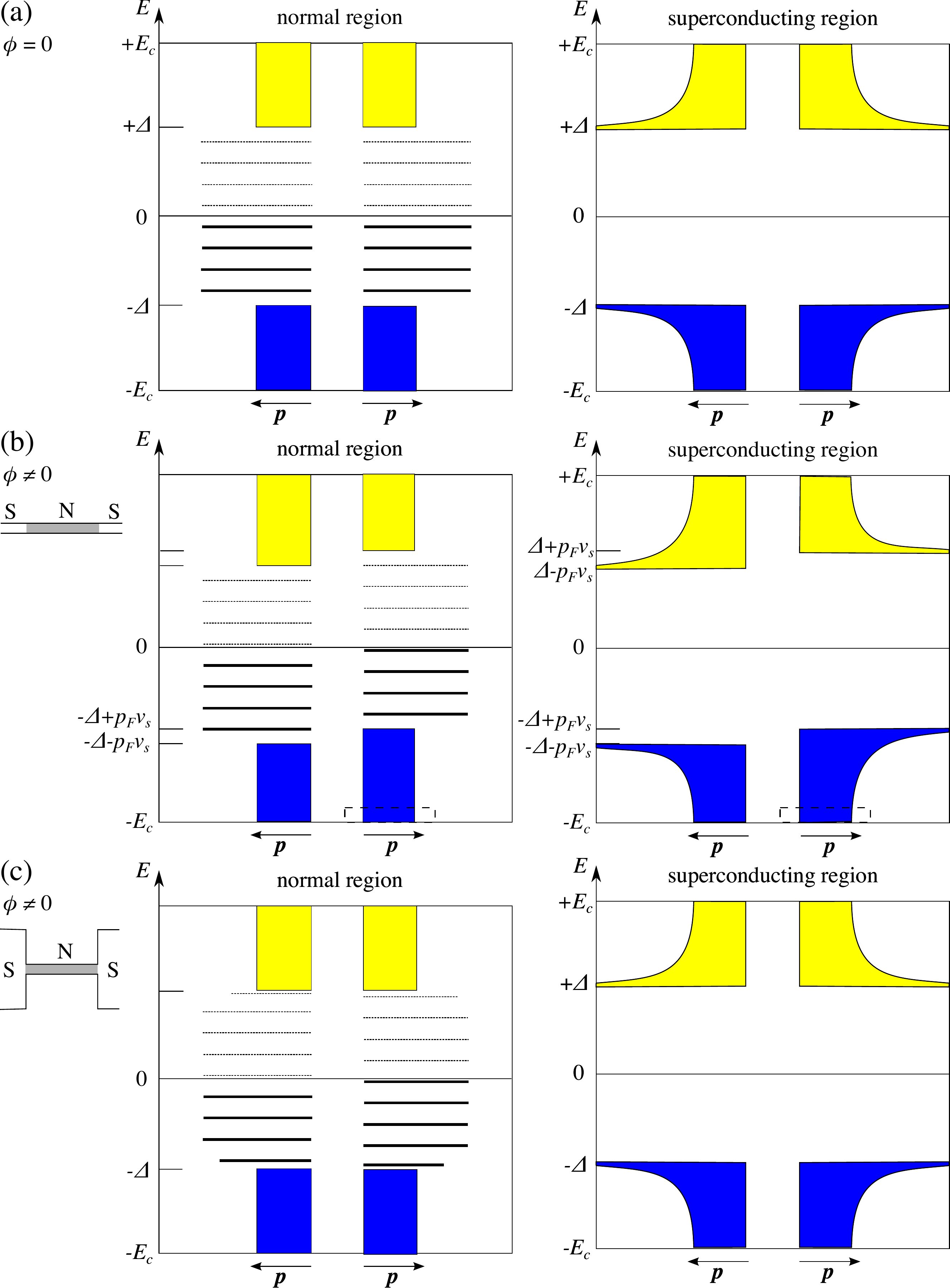} 
   \caption{The fermion energy spectra of normal (left) and superconducting (right) regions in a long ballistic SNS junction. The panels (a) represent equilibrium with zero phase difference, $\phi=0$, where the energy levels with momentum $\bm p$ to the right (right arrow) and to the left (left arrow) have equal energies. The lower panels (b-c) represent situation with a constant positive phase difference, (b) the case of uniform-thickness wire and (c) the case of the normal wire between two wider superconducting leads.   The levels forming continuum are shown with blue and yellow shading.  The width of the continuum levels in the superconducting region indicates the standard BCS density of levels. The discrete levels, which appear in the normal region, are denoted by horizontal lines. At $T=0$ levels with negative energy  are  filled and with positive energy are empty. The current arises from imbalance of the left and right momentum levels. The supercurrent corresponds to the net momentum imbalance indicated by the dashed rectangle in the right panel (b). With wide leads (c) the asymmetry in the superconducting region is negligible. 
   For more explanation see the main text and for mathematical formulation the Appendix. }
   \label{f.Jermionspectrum}
\end{figure}

Figure \ref{f.Jermionspectrum}(a) represents the equilibrium energy spectrum of the normal and superconductor parts of a ballistic SNS junction. The Fermi level is selected as zero of energy, and we only consider levels within the energy range from $-E_c$ to $+E_c$. The constant $E_c$ is arbitrary except that it is assumed to be much larger than the temperature or the energy gap ($E_c\gg k_BT$, $\Delta$), but much smaller than the Fermi energy ($E_c\ll E_F$). We consider a single quantum channel in the conductor, where the momentum is approximately $p=\pm p_F$, where $p_F$ is the channel dependent Fermi momentum.  In a superconductor there is no energy levels within the gap ($|E|<\Delta$) as they are pushed to energies $|E|>\Delta$, and cause the characteristic BCS density of levels \cite{BCS57}. In the normal part the levels within the gap have discrete energies due to repeated Andreev reflection at the NS interfaces \cite{Kulik69}. At low energies the level spacing $\delta=hv_F/2L$, where $v_F$ is the Fermi velocity.
The current carried by a single bound level is equal to the current-carrying capacity of the normal-state continuum levels in the energy interval equal to the level spacing.
When no phase difference is applied, the levels corresponding to the opposite momentum directions are degenerate. 
With equilibrium occupations, there is thus no current. 

The ``semiconductor picture'' shown in Fig.\ \ref{f.Jermionspectrum} is a double representation in the sense that the occupations $n(p,E)$ of the levels obey symmetry relation $n(p, E)=1-n(-p,-E)$. This means that only half of the levels shown are independent. We claim that keeping the double representation in general leads to a simpler picture than artificially removing half of the levels. In particular, keeping the negative energy levels allows simple interpretation of the supercurrent as filled negative energy levels. 

Figure \ref{f.Jermionspectrum}(b) illustrates a current carrying state in a uniform-thickness SNS wire, where the number of channels in the normal region and in the leads is the same. Here the supercurrent corresponds to the difference in the area of the two blue regions in the right panel. We see that large part of the areas cancel each other, but there are more levels with momentum to the right, which gives net supercurrent to the right. The difference is easiest to evaluate from the dashed rectangle, where the density of levels there is the same as in the normal state.

In the normal region both the continuum and discrete levels contribute to the current. In the case of uniform-thickness wire [Fig.\ \ref{f.Jermionspectrum}(b)], the levels in the normal region are shifted similarly as in the superconducting region. When the occupations of the levels are not changed, the current in the normal region corresponds to the dashed rectangle. This current is called {\em condensate current} according to the definition on the first page of Ref.\  \cite{Sonin21}.
The currents in the dashed rectangles in the normal and superconducting sides are equal, and thus the current is conserved. 

A slightly different case is shown in  Fig.\ \ref{f.Jermionspectrum}(c), where we consider an identical normal wire but  connected to wider superconducting leads. The current in the leads is divided into many more conducting channels than in the normal region. Therefore the phase gradient and the shift of the spectrum in the leads is negligible. In the normal region there is rearrangement of the levels, as the bound levels have moved relative to $\phi=0$ case  \cite{Kulik69}, but the gap edges remain unchanged. Thus the levels are shifted relative to the gap edge, and some new discrete levels may appear and some others disappear. The current with $T=0$ occupations in the normal layer is called {\em vacuum current}. Also this definition is in accordance with the one on first page of Sonin's paper \cite{Sonin21}, where the vacuum current is defined as the current in the normal wire with vanishing phase gradient in the leads. The wider leads introduced here is a trick to realize vacuum current as the only contribution to the current, which is impossible in a uniform-thickness wire.

We now claim that the vacuum current [Fig.\ \ref{f.Jermionspectrum}(c)] is identical to the condensate current [Fig.\ \ref{f.Jermionspectrum}(b)]. A simple principle that would give this result is that the current is fully determined by level energies and their occupations near the the Fermi surface, $|E|\ll \Delta$.  The justification for this could be that leaving some unoccupied momentum levels deeper inside the Fermi surface would not lead to a state with low energy, and the full occupation of these levels leads to cancelation of the contribution to current. Since the bound levels near the Fermi energy in the two cases, corresponding to Figs.\ \ref{f.Jermionspectrum}(b) and (c) are identical with the same phase difference $\phi$ \cite{Kulik69,Bardeen72}, this principle leads to equal currents in the two cases. 

Alternatively, the identity of the two currents can be shown by direct calculation. The condensate current can be obtained by simple Galilean invariance argument, as discussed by Bardeen and Johnson \cite{Bardeen72}. The vacuum current was first calculated by Ishii \cite{Ishii70}. We comment several aspects of his calculation. Firstly, he considered an unshifted superconducting gap, which means that his calculation directly addresses the vacuum current rather than the condensate current. Secondly, he considers a normal wire with many conducting channels, which appears as integration over transverse momentum. This is not a limitation since his results can trivially be generalized to arbitrary number of conducting channels by replacing the integral with a summation over channels with a proper coefficient \cite{Furusaki91}. An important result Ishii gets is the expression (3.5), which gives the current expressed as sum over the imaginary (Matsubara) frequencies. The same result in different forms has since been derived using different methods in Refs.\ \cite{Svidzinsky71,Svidzinsky73,Bezuglyi75,Kupriyanov81,Furusaki91,Golubov04}. One more derivation  is given in Appendix. It is also worth pointing out that this result is valid for any temperature and for any length $L$ of the normal region. Evaluation of the expression at $T=0$ for a long junction gives the vacuum current that is linear in $\phi$ in the range $|\phi|<\pi$ and agrees also in magnitude with the condensate current. 

Ishii further makes an analytic continuation from the imaginary frequencies to real frequencies in his expression for the vacuum current. Represented as integral over real frequencies, the current is seen to arises from two different contributions. One comes from the bound levels and the other from the continuum. 

The problem in Ref.\  \cite{Sonin21} is the the vacuum current has a different form from the condensate current. [To see this, compare Fig.\ 2(a)-(c) as function of $\theta_0$ with  Eq.\ (52) as function of $\theta_s$.] At least one source for this error is that the contribution to the vacuum current from the continuum levels is found to vanish in Ref.\  \cite{Sonin21}. 
As a consequence, the prediction of Ref.\  \cite{Sonin21} on shifted current-phase relation ($\pi$ or $\theta$ junction), as well as on the current that vanishes slowly with temperature, are incorrect.

Because of the same form of the condensate and vacuum currents, there is no need to consider them separately. Thus the division of the phase difference $\phi$ into the two contributions in  Ref.\  \cite{Sonin21} is not needed. 
The simplest procedure is to calculate the current of the Andreev levels for a given phase difference. This then determines the current, which then determines the phase gradient in the superconducting leads. In order to properly treat the current conversion at the N-S interfaces, a simple step potential model has to be replaced by self-consistent calculation of the order parameter profile, as for example in Ref.\ \cite{Riedel96}. The conclusion is that this has only minor effect on the low-energy levels discussed above.

I thank Vadim Geshkenbein, Andrei Shelankov and Edouard Sonin for commenting the manuscript.

\appendix*
\section{}

This appendix includes derivation of Ishii's result (3.5) \cite{Ishii70} based on Eilenberger equation and mathematical basis of Fig.\ \ref{f.Jermionspectrum}. The notation is close to that of Ref.\ \cite{Serene83}. For simplicity we neglect the effects of Fermi-liquid interactions. It would be possible to take them into account at the expense of slightly more complicated equations.

The Eilenberger equation for the $2\times 2$ matrix quasiclassical propagator $\widehat g$ is
\begin{equation} 
[{\rm i}\epsilon_n
\widehat\tau_3
-\widehat\Delta({\bm R}), 
\widehat g(\hat{\bm p},{\bm R};\epsilon_n)]+{\rm i}\hbar{v}_{\rm
F}\hat{\bm p}\cdot\bm\nabla_{\bm R}\widehat g(\hat{\bm p},{\bm R};\epsilon_n)=0 
\label{e.eil}\end{equation}
and the normalization condition
\begin{equation}  
\widehat g(\hat{\bm p},{\bm R};\epsilon_n)\widehat g(\hat{\bm p},{\bm R};\epsilon_n)=-\pi^2. 
\label{e.norm}\end{equation}
Here $\bm R$ is the location, $\hat{\bm p}$ the momentum direction, $\epsilon_n=\pi T(2n+1)$  the Matsubara energies at temperature $T$ with integer $n$, $v_{\rm F}$ is the Fermi velocity, $[\widehat a,\widehat b]=\widehat a\widehat b- \widehat b \widehat a$, and  $\widehat\tau_1$, $\widehat\tau_2$, and $\widehat\tau_3$ are the Pauli matrices. The gap matrix $\widehat\Delta={\rm i}(\Delta_{\rm i}\widehat\tau_1+\Delta_{\rm
r}\widehat\tau_2)$, where $\Delta_{\rm r}$ and $\Delta_{\rm i}$ are the real and imaginary parts of the complex gap $\Delta=\Delta_{\rm r}+{\rm i} \Delta_{\rm i}$.

The homogeneous solution of (\ref{e.eil}) and (\ref{e.norm}) is
\begin{equation}
\widehat g_b(\epsilon_n)
=\frac{\pi} 
{\sqrt{\epsilon_n^2+|\Delta|^2}}\left(\begin{array}{cc}-{\rm i}\epsilon_n&\Delta\\ -\Delta^*&{\rm i}\epsilon_n\end{array}\right).
\label{e.bulkg} \end{equation}
The solution in the presence of phase gradient $\Delta(\bm R)=\Delta_0e^{i\bm q\cdot\bm R}$ is 
\begin{equation}
\widehat g(\hat{\bm p},{\bm R};\epsilon_n)
=\frac{\pi} 
{\sqrt{(\epsilon_n+{\rm i}a)^2+|\Delta|^2}}\left(\begin{array}{cc}-{\rm i}(\epsilon_n+{\rm i}a)&\Delta(\bm R)\\ -\Delta^*(\bm R)&{\rm i}(\epsilon_n+{\rm i}a)\end{array}\right)
\label{e.bulkgflow} \end{equation}
with $a=mv_{\rm F}\hat{\bm p}\cdot\bm v_s$, where the superfluid velocity $\bm v_s =(\hbar/2m)\bm q$ and $m$ is the particle mass.

The Eilenberger equation (\ref{e.eil}) is a set of first order differential equations  along trajectories, which are lines parallel to $\hat{\bm p}$.
For a constant $\Delta$ it has two exponential solutions (\ref{e.eil}) \begin{eqnarray}
\widehat g_\pm
=\left(\begin{array}{cc}{\rm i}|\Delta|^2&\Delta(\mp\sqrt{\epsilon_n^2+|\Delta|^2}+\epsilon_n) \\
 \Delta^*(\mp\sqrt{\epsilon_n^2+|\Delta|^2}-\epsilon_n)&-{\rm i}|\Delta|^2\end{array}\right)
 \exp\left(\frac{\pm 2\sqrt{\epsilon_n^2+|\Delta|^2}u}{\hbar v_{\rm F}}
\right),
\label{e.exponential} \end{eqnarray}
where  $u$ is the parameter along a trajectory, $\bm R(u)=\bm R(0)+\hat{\bm p}u$.

For SNS junction consider a trajectory where $\Delta=\Delta_0 e^{-{\rm i}\phi/2}$ for $u<-L/2$, $\Delta=0$ for $|u|<L/2$, and $\Delta=\Delta_0 e^{{\rm i}\phi/2}$ for $u>L/2$. To simplify the formulas we take  $\Delta_0$ real and use $\alpha=\sqrt{\epsilon_n^2+\Delta_0^2}$. In region $u<-L/2$ the propagator is a linear combination of the bulk and growing solution,
\begin{eqnarray}
\frac{\widehat g(u<-L/2)}{\pi}=\frac{1} 
{\alpha}\left(\begin{array}{cc}-{\rm i}\epsilon_n&\Delta_0 e^{-{\rm i}\phi/2}\\ -\Delta_0 e^{{\rm i}\phi/2}&{\rm i}\epsilon_n\end{array}\right)
+A\left(\begin{array}{cc}{\rm i}\Delta_0&(-\alpha+\epsilon_n) e^{-{\rm i}\phi/2}\\
  (-\alpha-\epsilon_n)e^{{\rm i}\phi/2}&-{\rm i}\Delta_0\end{array}\right)
 \exp\left(\frac{ 2\alpha u}{\hbar v_{\rm F}}\right).
\label{e.leftee} \end{eqnarray}
In region $u>L/2$ the propagator is a linear combination of the bulk and decaying solution,
\begin{eqnarray}
\frac{\widehat g(u>L/2)}{\pi}=\frac{1} 
{\alpha}\left(\begin{array}{cc}-{\rm i}\epsilon_n&\Delta_0 e^{{\rm i}\phi/2}\\ -\Delta_0 e^{-{\rm i}\phi/2}&{\rm i}\epsilon_n\end{array}\right)
+B\left(\begin{array}{cc}{\rm i}\Delta_0&(\alpha+\epsilon_n) e^{{\rm i}\phi/2}\\
  (\alpha-\epsilon_n)e^{-{\rm i}\phi/2}&-{\rm i}\Delta_0\end{array}\right)
 \exp\left(-\frac{2\alpha u}{\hbar v_{\rm F}}\right).
\label{e.rightee} \end{eqnarray}
In the middle region it is most convenient to look for the propagator in the form
\begin{eqnarray}
\frac{\widehat g(|u|<L/2)}{\pi}= 
C\left(\begin{array}{cc}1&0\\ 0&-1\end{array}\right)
+D\left(\begin{array}{cc}0&0\\
1&0\end{array}\right)
 \exp\left(\frac{2\epsilon_nu}{\hbar v_{\rm F}}
\right)+E\left(\begin{array}{cc}0&1\\
0&0\end{array}\right)
 \exp\left(-\frac{2\epsilon_nu}{\hbar v_{\rm F}}\right).
\label{e.midee} \end{eqnarray}
We should now require the continuity of $\widehat g$ at $u=- L/2$ and $u=L/2$. This fixes the coefficients $A$, $B$, $C$, $D$ and $E$. We get in the middle region the propagator
\begin{eqnarray}
\frac{ \widehat g(|u|<L/2)}{\pi}=\frac{1}{(\alpha+\epsilon_n)e^z+(\alpha-\epsilon_n)e^{-z}}
\left(\begin{array}{cc}-{\rm i}[(\alpha+\epsilon_n)e^z-(\alpha-\epsilon_n)e^{-z}]&2\Delta_0\exp(-2\epsilon_nu/\hbar v_{\rm F})\\ 
-2\Delta_0\exp(2\epsilon_nu/\hbar v_{\rm F})&{\rm i}[(\alpha+\epsilon_n)e^z-(\alpha-\epsilon_n)e^{-z}]\end{array}\right),
\label{e.figmidle}\end{eqnarray}
where $z=\epsilon_nL/\hbar v_F+{\rm i}\phi/2$. The obtained $\widehat g$ obeys the general symmetry relations, e.g. $g_{11}(\hat{\bm p},{\bm R};-\epsilon_n)=g_{11}^*(\hat{\bm p},{\bm R};\epsilon_n)$ and $g_{11}(-\hat{\bm p},{\bm R};-\epsilon_n)=g_{22}(\hat{\bm p},{\bm R};\epsilon_n)=-g_{11}(\hat{\bm p},{\bm R};\epsilon_n)$. 

The equilibrium current density $\bm j$ in bulk can be calculated using
\begin{eqnarray}
\bm j(\bm R)= 2ev_FN(0)T\sum_{n=-\infty}^\infty\int\frac{d\Omega_p}{4\pi}\hat{\bm p}\,g_{11}(\hat{\bm p},\bm R,\epsilon_n),
\label{e.SR2.12c}\end{eqnarray}
where $\int d\Omega_p=\int_0^{2\pi}d\phi_p\int_0^\pi d\theta_p\sin\theta_p$ is the integral over the solid angle formed by all $\hat{\bm p}=(\hat{\bm x}\cos\phi_p+\hat{\bm y}\sin\phi_p)\sin\theta_p+\hat{\bm z}\cos\theta_p$, $2N(0)=mp_{\rm F}/\pi^2\hbar^3$ is the density of levels at the Fermi surface, and $e$ is the electron charge.
In order to calculate the current in a channel, this formula can be transformed to the form
\begin{eqnarray}
J=\frac{e}{\pi\hbar}T\sum_{n=-\infty}^\infty \sum_{i=1}^M \left[g_{11}(i,\hat p_z>0;\epsilon_n)-g_{11}(i,\hat p_z<0;\epsilon_n)\right].
\label{e.curscha}\end{eqnarray}
Here $i$ is the index labeling the $M$ open channels and $\hat p_z$ is the component of momentum direction along the channel. Using this and (\ref{e.figmidle}) we get
\begin{eqnarray}
J=\frac{2e}{\hbar}T\sum_{n=-\infty}^\infty \sum_{i=1}^M\frac{\Delta^2\sin\phi}{(2\epsilon_n^2+\Delta^2)\cosh(2\epsilon_nL/v_{Fi})+
2\epsilon_n\sqrt{\epsilon_n^2+\Delta^2}\sinh(2\epsilon_nL/v_{Fi})+ \Delta^2\cos\phi}
\label{e.ishimf}\end{eqnarray}
This result is equivalent to Ishii's formula (3.5) \cite{Ishii70}, and has also been obtained in Refs.\ \cite{Svidzinsky71,Svidzinsky73,Bezuglyi75,Kupriyanov81,Furusaki91,Golubov04}.
Based on this, Ishii derived for a long junction at $T=0$ the current
\begin{eqnarray}
J=\sum_{i=1}^M\frac{ev_{Fi}}{\pi L}\phi \quad {\rm for}\ |\phi|<\pi
\label{e.ishimfd0}\end{eqnarray}
and repeated periodically with period $2\pi$. (Note the correction to Ref.\ \cite{Ishii70} pointed out in Ref.\ \cite{Ishii72}.) At general temperature the current can be evaluated numerically. The special case of a SS point contact is obtained in the case of $L=0$. The current in this case can be solved analytically at all temperatures \cite{Kulik77}.

The Matsubara technique is a highly convenient tool to calculate the equilibrium currents in superconductors. The problem is that it is a rather abstract theory. In order to gain physical understanding, one has to do analytic continuation from $\epsilon=i\epsilon_n$ to the real axis in the complex $\epsilon$ plane \cite{FW}. As an example, consider the real-frequency form of the current formula (\ref{e.SR2.12c}) when applied to a bulk superconductor described by the propagator (\ref{e.bulkgflow}), 
\begin{eqnarray}
\bm j(\bm R,t)=ev_FN(0)\int\frac{d\Omega_p}{4\pi}\hat{\bm p}
\int_{-E_c}^{E_c} d\epsilon\,\frac{|\tilde\epsilon|}{\sqrt{\tilde\epsilon^2-|\Delta|^2}}\theta(\tilde\epsilon^2-|\Delta|^2)\left[\phi_{B1}(\hat{\bm p},{\bm R};\epsilon,t)+\phi_{B2}(\hat{\bm p},{\bm R};\epsilon,t)\right].
\label{e.jpb12}\end{eqnarray}
Here $\tilde\epsilon=\epsilon-a$, $\theta(x)$ is the Heaviside step function, and $\phi_{B1}$ and $\phi_{B2}$ are the symmetrized distribution functions for particle and hole type excitations. The analytic continuation from (\ref{e.SR2.12c}) gives that the distributions are equal to the symmetrized Fermi functions, $\phi_{B1}(\epsilon)=\phi_{B2}(\epsilon)=-\frac12\tanh(\epsilon/2T)=1/(e^{\epsilon/T}+1)-\frac12$, but a more general derivation (section 7 of Ref.\ \cite{Serene83}) shows that formula (\ref{e.jpb12}) is valid also for non-equilibrium quasiparticle  distributions. By definition, for particle-type excitations the propagation direction is the same as the momentum direction, but for hole-type excitations the two directions are opposite to each other.

Equation  (\ref{e.jpb12}) now allows a simple interpretation of the current in superconductor, as explained in the main text and in the right hand side panels of Fig.\ \ref{f.Jermionspectrum}. The factor $\theta(\tilde\epsilon^2-|\Delta|^2)$ indicates that there are no levels within the gap.  The factor $|\tilde\epsilon|/\sqrt{\tilde\epsilon^2-|\Delta|^2}$ gives the BCS density of levels.  The current at $T=0$ arises from imbalance of opposite momentum directions of the filled negative energy levels. For $a=0$, there is no imbalance and the current vanishes [Fig.\ \ref{f.Jermionspectrum}(a)].
For $a\not=0$, there is imbalance and non-vanishing current [Fig.\ \ref{f.Jermionspectrum}(b)]. Since the difference between particle and hole type excitations is not important for electric current, they are not shown separately in Fig.\ \ref{f.Jermionspectrum}.

In order to obtain similar understanding of SNS junction, one has to continue (\ref{e.ishimf}) to the real $\epsilon$ axis. The zeros of the denominator give the energies of the bound levels \cite{Kulik69}. In general, the current can be interpreted to arise from imbalance of both the bound and the continuum levels \cite{Ishii70}. For a SS point contact ($L=0$) the bound state energies are $\epsilon=\pm\Delta_0\cos(\phi/2)$ and they are solely responsible for the current \cite{Kulik77,Thuneberg06}.

\end{document}